\documentclass{lncse}
\usepackage{multicol}
\usepackage{makeidx}
\usepackage{epsfig}
\usepackage{amssymb,amsmath}
\usepackage{citesort}
\usepackage{epic}
\usepackage{eepic}

\usepackage{amsmath}

\usepackage{fancyheadings}
\usepackage{graphicx}
\usepackage{array}

\newcommand{\figref}[1]{Fig.~\ref{#1}}

\newcommand{\Figref}[1]{Fig.~\ref{#1}}

\newcommand{\psinpw}[2]{\epsfig{file=#1.ps,width=#2}}

\begin{document}

\frontmatter          

\pagestyle{headings}  
\addtocmark{CA - Applications} 

\title{Cellular Automaton Approach to Pedestrian Dynamics - Applications}

\author{Carsten Burstedde, Ansgar Kirchner, Kai Klauck, Andreas 
Schadschneider \and Johannes Zittartz}

\institute{Institut f\"ur Theoretische Physik, Universit\"at zu K\"oln\\
  D-50937 K\"oln, Germany\\
  email: cb,aki,kok,as,zitt@thp.uni-koeln.de
}

\maketitle              


\begin{abstract}
We present applications of a cellular automaton approach
to pedestrian dynamics introduced in \cite{part1,ourpaper}.
It is shown that the model is able to reproduce collective effects
and self-organization phenomena encountered in pedestrian traffic,
e.g.\ lane formation in counterflow through a large corridor and
oscillations at doors. Furthermore we present simple examples
where the model is applied to the simulation of evacuation processes.
\end{abstract}


\section{Introduction}
\label{sec_intro}

In Part I \cite{part1} (see also \cite{ourpaper}) we have introduced
a two-dimensional cellular automaton model for the description of
pedestrian dynamics. In the model the space is discretized into
small cells which can either be empty or occupied by exactly one
pedestrian. These can move to one of the neighbouring cells where
the direction of motion is chosen with certain probabilities.
These probabilities depend on the
\begin{itemize}
\item[1.] occupation of the targets cells: Motion to a cell already
occupied is forbidden.
\item[2.] matrix of preference: It encodes the direction of motion
including the average velocity of the pedestrian.
\item[3.] value of the dynamic floor field: This corresponds to
a virtual trace left by the other pedestrians.
\item[4.] value of the static floor field: This allows to specify
preferred regions, e.g.\ to incorporate effects of the geometry
of a building etc.
\end{itemize}
The introduction of the floor fields allows to translate the long-ranged
interactions between pedestrians into  local ones. Thus the model
is extremely efficient in computer simulations and large crowds can
be simulated much faster than real time. Despite this simplicity
the cellular automaton allows to reproduce the collective effects
and self-organization phenomena encountered in pedestrian traffic.
Examples will be given in the following sections.


\section{Self-Organization Phenomena}
\label{selforg}

One of the major design principles of our model is to provide the
simulated individuals with as little intelligence as possible.  
There have been approaches\footnote{For a more complete discussion
of other modelling approaches, see \cite{part1} and references
therein.} where the pedestrians examine their
surroundings or look ahead a certain amount of grid cells in order to
decide where and how far to move.  It has also been proposed to select
a target cell and repeatedly choose a different one if it is found
occupied. 

These approaches imply the need for more complicated algorithms which
reduce the simulation speed.  Furthermore, it inhibits the 
unambiguous definition of the update procedure used.  
We avoid this strategy and rely on the concept of self-organization
to model pedestrian behaviour.

We argue that simple, local update rules for the pedestrians and
optionally one or more floor fields are sufficient to yield a richness
of complex phenomena.  Obviously this route is superiour concerning
the computational efficiency and even allows for faster-than-real-time
simulations of large crowds, e.g.\ in evacuation processes in public
buildings.

\subsection{Lane Formation}
\label{laneformation}

The most prominent collective phenomenon is the formation of lanes out
of an unordered group of pedestrians \cite{social}.  This corresponds to a
spontaneous breaking of the symmetry of the particle number
distribution in space.
Our simulations show that an even as well as an odd number of lanes 
may be formed. The latter corresponds to a spontaneous breaking of 
the left-right symmetry of the system.

\begin{figure}
  \centerline{\psinpw{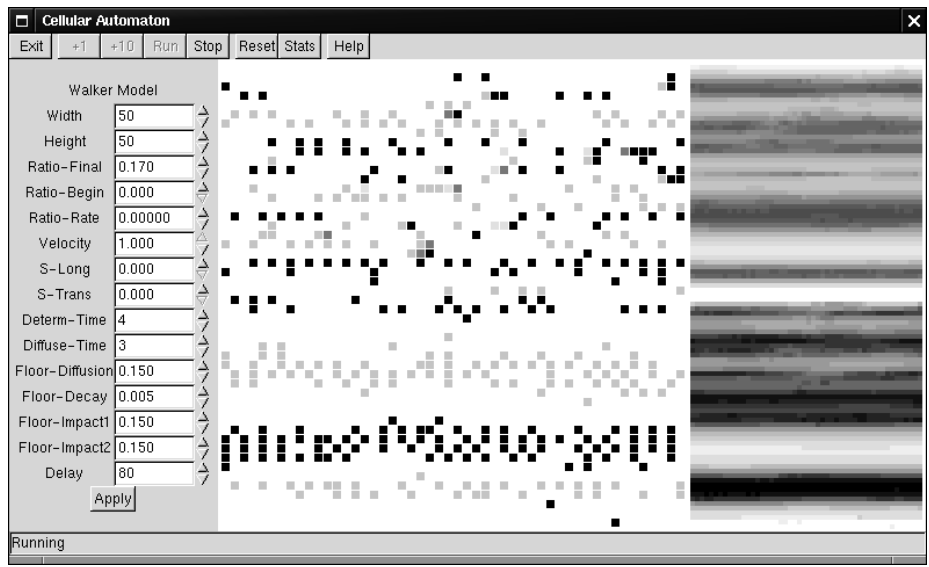}{\linewidth}}
  \caption{Snapshot of a simulation with $\rho = 0.17$, $w = h =
    50$. The left part shows the parameter control. The central window
    is the corridor and the light and dark squares are right- and
    left-moving pedestrians, respectively. The right part shows the
    floor fields for the two species.}
  \label{snap1}
\end{figure}

We present simulations of a rectangular corridor which is populated by
two species of pedestrians moving in opposite directions (see
\figref{snap1}). Parallel to the direction of motion we assume the
existence of walls.  Orthogonal to the direction of motion we
investigated both periodic and open boundary conditions.  The length
of the corridor is set to 200 cells.  Widths of 15 and 25 cells have
been used.

With periodic boundary conditions, the density of pedestrians is fixed
for each run.  It is ensured that the overall number of
pedestrians is evenly divided by the numbers for the different
species: with two species, one is moving to the left and the other to
the right.  For open boundaries, we fix the rate at which pedestrians
enter the system at the boundaries.  The
pedestrians leave the system as soon as they reach the opposite end of
the corridor.

\Figref{snap1} shows the graphical frontend running a simulation of
a small periodic system.  The lanes can be spotted easily, both in the
main window showing the cell contents and the small windows on the
right showing the floor field intensity for the two species.

This model clearly provides the option for a complete jam.  The
jamming probability with periodic boundaries at constant density
increases with the length of the system.  An open system can be
thought of as the limit of an infinitely long periodic system,
although density and entry rates do not correspond absolutely.

We performed several runs for different densities and insertion rates,
respectively.  The focus of our attention is the parameter range where
the transition from a stable flow to a complete jam takes place.

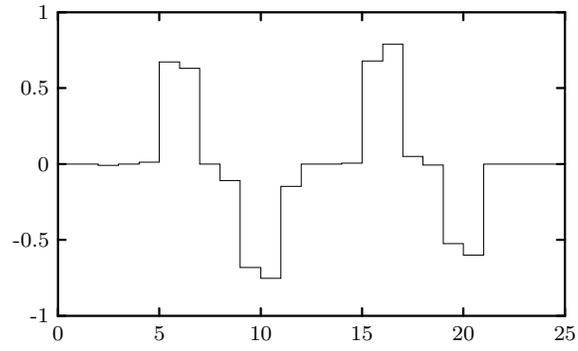
\begin{figure}
  \centerline{
\setlength{\unitlength}{0.15pt}
\begin{picture}(1279,836)(154,20)
\footnotesize
\thicklines \path(154,90)(174,90)
\thicklines \path(1433,90)(1413,90)
\put(132,90){\makebox(0,0)[r]{-1}}
\thicklines \path(154,282)(174,282)
\thicklines \path(1433,282)(1413,282)
\put(132,282){\makebox(0,0)[r]{-0.5}}
\thicklines \path(154,473)(174,473)
\thicklines \path(1433,473)(1413,473)
\put(132,473){\makebox(0,0)[r]{0}}
\thicklines \path(154,665)(174,665)
\thicklines \path(1433,665)(1413,665)
\put(132,665){\makebox(0,0)[r]{0.5}}
\thicklines \path(154,856)(174,856)
\thicklines \path(1433,856)(1413,856)
\put(132,856){\makebox(0,0)[r]{1}}
\thicklines \path(154,90)(154,110)
\thicklines \path(154,856)(154,836)
\put(154,45){\makebox(0,0){0}}
\thicklines \path(410,90)(410,110)
\thicklines \path(410,856)(410,836)
\put(410,45){\makebox(0,0){5}}
\thicklines \path(666,90)(666,110)
\thicklines \path(666,856)(666,836)
\put(666,45){\makebox(0,0){10}}
\thicklines \path(921,90)(921,110)
\thicklines \path(921,856)(921,836)
\put(921,45){\makebox(0,0){15}}
\thicklines \path(1177,90)(1177,110)
\thicklines \path(1177,856)(1177,836)
\put(1177,45){\makebox(0,0){20}}
\thicklines \path(1433,90)(1433,110)
\thicklines \path(1433,856)(1433,836)
\put(1433,45){\makebox(0,0){25}}
\thicklines \path(154,90)(1433,90)(1433,856)(154,856)(154,90)

\thinlines \path(154,473)(154,473)(205,473)(205,473)(256,473)(256,470)(307,470)(307,473)(359,473)(359,478)(410,478)(410,731)(461,731)(461,715)(512,715)(512,473)(563,473)(563,431)(614,431)(614,212)(666,212)(666,185)(717,185)(717,417)(768,417)(768,473)(819,473)(819,473)(870,473)(870,476)(921,476)(921,734)(973,734)(973,776)(1024,776)(1024,493)(1075,493)(1075,471)(1126,471)(1126,272)(1177,272)(1177,244)(1228,244)(1228,473)(1280,473)(1280,473)(1331,473)(1331,473)(1382,473)(1382,473)(1433,473)

\end{picture}}
  \caption{Velocity profile of a periodic system with $\rho =
    0.12$.}
  \label{H25R010}
\end{figure}

\begin{figure}
  \centerline{
\setlength{\unitlength}{0.15pt}
\begin{picture}(1279,836)(154,20)
\footnotesize
\thicklines \path(154,167)(174,167)
\thicklines \path(1433,167)(1413,167)
\put(132,167){\makebox(0,0)[r]{-0.4}}
\thicklines \path(154,320)(174,320)
\thicklines \path(1433,320)(1413,320)
\put(132,320){\makebox(0,0)[r]{-0.2}}
\thicklines \path(154,473)(174,473)
\thicklines \path(1433,473)(1413,473)
\put(132,473){\makebox(0,0)[r]{0}}
\thicklines \path(154,626)(174,626)
\thicklines \path(1433,626)(1413,626)
\put(132,626){\makebox(0,0)[r]{0.2}}
\thicklines \path(154,779)(174,779)
\thicklines \path(1433,779)(1413,779)
\put(132,779){\makebox(0,0)[r]{0.4}}
\thicklines \path(154,90)(154,110)
\thicklines \path(154,856)(154,836)
\put(154,45){\makebox(0,0){0}}
\thicklines \path(410,90)(410,110)
\thicklines \path(410,856)(410,836)
\put(410,45){\makebox(0,0){5}}
\thicklines \path(666,90)(666,110)
\thicklines \path(666,856)(666,836)
\put(666,45){\makebox(0,0){10}}
\thicklines \path(921,90)(921,110)
\thicklines \path(921,856)(921,836)
\put(921,45){\makebox(0,0){15}}
\thicklines \path(1177,90)(1177,110)
\thicklines \path(1177,856)(1177,836)
\put(1177,45){\makebox(0,0){20}}
\thicklines \path(1433,90)(1433,110)
\thicklines \path(1433,856)(1433,836)
\put(1433,45){\makebox(0,0){25}}
\thicklines \path(154,90)(1433,90)(1433,856)(154,856)(154,90)


\thinlines \path(154,484)(154,484)(205,484)(205,416)(256,416)(256,544)(307,544)(307,411)(359,411)(359,423)(410,423)(410,631)(461,631)(461,468)(512,468)(512,490)(563,490)(563,336)(614,336)(614,523)(666,523)(666,655)(717,655)(717,552)(768,552)(768,439)(819,439)(819,404)(870,404)(870,346)(921,346)(921,404)(973,404)(973,596)(1024,596)(1024,645)(1075,645)(1075,429)(1126,429)(1126,397)(1177,397)(1177,377)(1228,377)(1228,563)(1280,563)(1280,591)(1331,591)(1331,531)(1382,531)(1382,454)(1433,454)

\end{picture}}
  \caption{Velocity profile of an open system at $x =
    \frac{L}{2}$ with $\alpha = 0.04$.}
  \label{H25A004R0}
\end{figure}
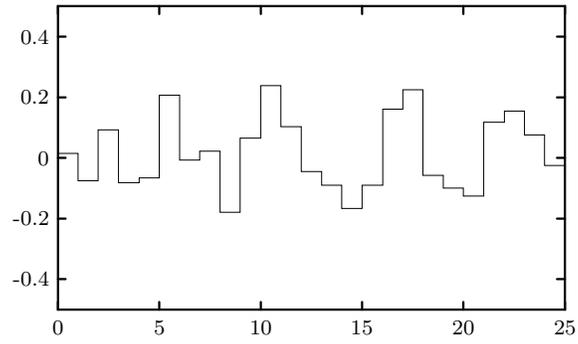

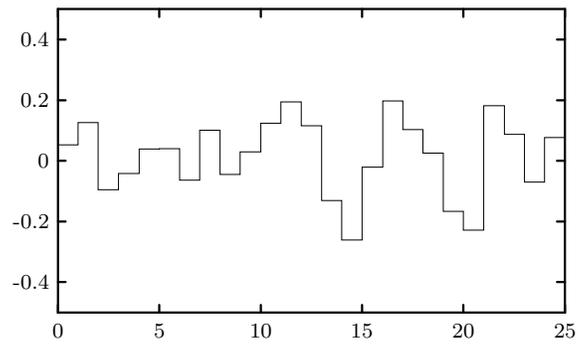
\begin{figure}
  \centerline{
\setlength{\unitlength}{0.15pt}
\begin{picture}(1279,836)(154,20)
\footnotesize
\thicklines \path(154,167)(174,167)
\thicklines \path(1433,167)(1413,167)
\put(132,167){\makebox(0,0)[r]{-0.4}}
\thicklines \path(154,320)(174,320)
\thicklines \path(1433,320)(1413,320)
\put(132,320){\makebox(0,0)[r]{-0.2}}
\thicklines \path(154,473)(174,473)
\thicklines \path(1433,473)(1413,473)
\put(132,473){\makebox(0,0)[r]{0}}
\thicklines \path(154,626)(174,626)
\thicklines \path(1433,626)(1413,626)
\put(132,626){\makebox(0,0)[r]{0.2}}
\thicklines \path(154,779)(174,779)
\thicklines \path(1433,779)(1413,779)
\put(132,779){\makebox(0,0)[r]{0.4}}
\thicklines \path(154,90)(154,110)
\thicklines \path(154,856)(154,836)
\put(154,45){\makebox(0,0){0}}
\thicklines \path(410,90)(410,110)
\thicklines \path(410,856)(410,836)
\put(410,45){\makebox(0,0){5}}
\thicklines \path(666,90)(666,110)
\thicklines \path(666,856)(666,836)
\put(666,45){\makebox(0,0){10}}
\thicklines \path(921,90)(921,110)
\thicklines \path(921,856)(921,836)
\put(921,45){\makebox(0,0){15}}
\thicklines \path(1177,90)(1177,110)
\thicklines \path(1177,856)(1177,836)
\put(1177,45){\makebox(0,0){20}}
\thicklines \path(1433,90)(1433,110)
\thicklines \path(1433,856)(1433,836)
\put(1433,45){\makebox(0,0){25}}
\thicklines \path(154,90)(1433,90)(1433,856)(154,856)(154,90)


\thinlines \path(154,513)(154,513)(205,513)(205,570)(256,570)(256,400)(307,400)(307,441)(359,441)(359,503)(410,503)(410,504)(461,504)(461,424)(512,424)(512,550)(563,550)(563,439)(614,439)(614,496)(666,496)(666,568)(717,568)(717,622)(768,622)(768,562)(819,562)(819,373)(870,373)(870,273)(921,273)(921,457)(973,457)(973,624)(1024,624)(1024,552)(1075,552)(1075,493)(1126,493)(1126,346)(1177,346)(1177,298)(1228,298)(1228,612)(1280,612)(1280,540)(1331,540)(1331,419)(1382,419)(1382,532)(1433,532)

\end{picture}}
  \caption{Velocity profile of an open system at $x =
    \frac{L}{4}$ with $\alpha = 0.04$.}
  \label{H25A004R1}
\end{figure}

To obtain information about the lanes we measured the pedestrian
velocities at a cross section perpendicular to the direction of flow.
Selected velocity profiles are shown in
Figs.~\ref{H25R010}--\ref{H25A004R1}.

It is obvious that the lane formation in the periodic system works far
better than in the open system.  The floor field leads to an effective
attraction of identical pedestrians while different pedestrian species
separate.  This results in the formation of a stable pattern in the
periodic case.
\begin{figure}
  \centerline{\input{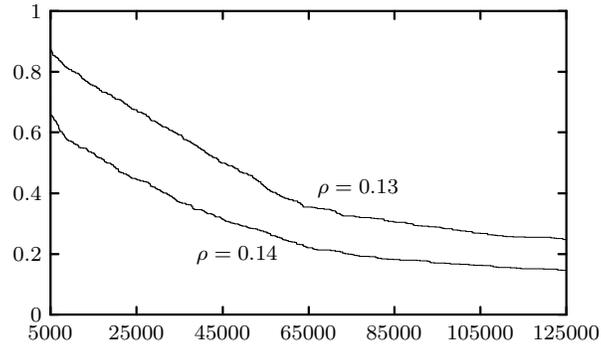}}
  \caption{The fraction of systems with nonzero flow versus the
    simulation time for two densities $0.13$ and $0.14$.}
  \label{timedevel}
\end{figure}

In a certain density regime, these lanes are metastable.  Spontaneous
fluctuations can disrupt the flow in one lane causing the pedestrians
to spread and interfere with other lanes.  Eventually the system can
run into a jam by this mechanism.  The average time after which the
system is blocked by a jam is an interesting observable which depends
on the density of pedestrians.  We observe large fluctuations of this
quantity which require many samples to find statistically
significant information.

To create \figref{timedevel} we sampled 700 initial conditions for
each of two densities and let them evolve for at most 125,000 timesteps.  
The diagram shows how the percentage of systems with nonzero
flow (i.e.\ jam-free systems) decreases with time.  In this regime,
the density has a big impact on this quantity.


\subsection{Different geometries}
\label{geovar}

Our model can easily be generalized for the use with different
geometrical setups.  The transition rules for the particles do not
need to be altered, and the computational speed does not suffer.

\begin{figure}
  \centerline{\psinpw{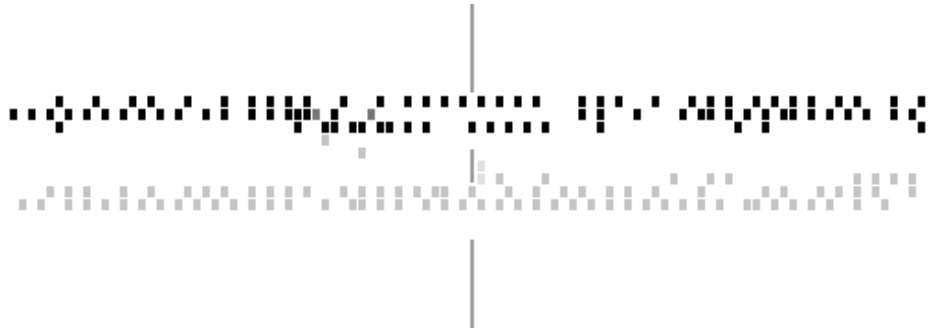}{\linewidth}}
  \caption{Separation of two species of opposite walking directions
    into two separate lanes catalyzed by two doors.}
  \label{doorsep}
\end{figure}

In \figref{doorsep} a wall with two doors has been added to the
system.  It is oriented perpendicular to the direction of flow. With
the appropriate choice of the doors' width and the distance between
them, they support the separation of the two particle species.

\begin{figure}
  \psinpw{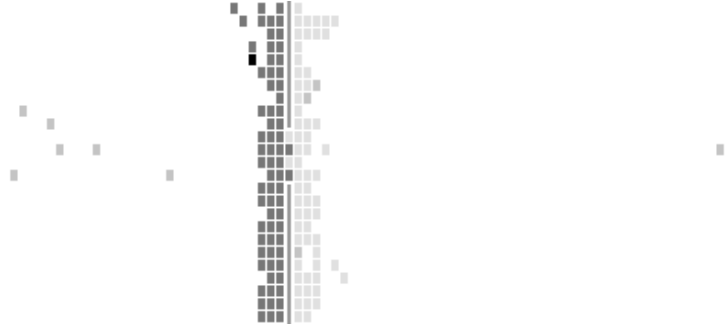}{\linewidth}
  \caption{Oscillations of the direction of flow: A group of particles
    of the same species break through a blockade at a door.}
  \label{oszis}
\end{figure}

Another feature of pedestrian behaviour are oscillations of the
direction of flow when two groups of opposite walking direction are
facing each other at a tight spot like a narrow door \cite{social}.
If we simulate this setup using our model, we can observe oscillation
on two time scales (provided that the parameters are suitably chosen):
An exchange between a blocked situation and a flow in both directions
is the main result.  Inside the blocked period small groups of only
one species can break through.  This is illustrated in \figref{oszis}.
These breaks can alternate between the two species and can therefore
be interpreted as oscillations.

\begin{figure}
  \centerline{\setlength{\unitlength}{0.0004in}
\begingroup\makeatletter\ifx\SetFigFont\undefined%
\gdef\SetFigFont#1#2#3#4#5{%
}
\fi\endgroup%
{\renewcommand{\dashlinestretch}{30}
\begin{picture}(4800,4800)(22,22)
\thicklines
\path(1822,4822)(22,4822)(22,3022)
\path(22,1822)(22,22)(1822,22)
\path(3022,22)(4822,22)(4822,1822)
\path(4822,3022)(4822,4822)(3022,4822)
\thinlines
\path(2047,4222)(2047,622)
\path(2197,622)(2197,4222)
\path(2647,4222)(2647,622)
\path(2797,622)(2797,4222)
\path(1822,847)(2122,547)(2422,847)
\path(3022,3997)(2722,4297)(2422,3997)
\path(622,2047)(4222,2047)
\path(622,2197)(4222,2197)
\path(622,2647)(4222,2647)
\path(622,2797)(4222,2797)
\path(3997,3022)(4297,2722)(3997,2422)
\path(847,2422)(547,2122)(847,1822)
\texture{55888888 88555555 5522a222 a2555555 55888888 88555555 552a2a2a 2a555555
	55888888 88555555 55a222a2 22555555 55888888 88555555 552a2a2a 2a555555 
	55888888 88555555 5522a222 a2555555 55888888 88555555 552a2a2a 2a555555 
	55888888 88555555 55a222a2 22555555 55888888 88555555 552a2a2a 2a555555 }
\put(2422,2422){\shade\ellipse{1006}{1006}}
\put(2422,2422){\ellipse{1006}{1006}}
\end{picture}
}}
  \caption{Typical flow pattern which emerges at a crossing.}
  \label{crossing}
\end{figure}
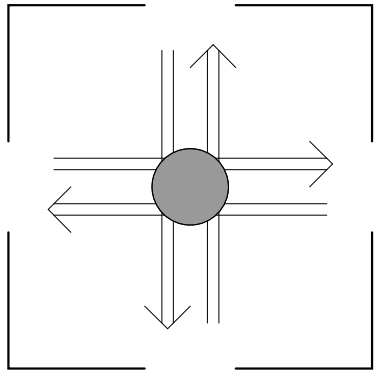

To simulate a crossing, four species of particles are inserted into
the system.  Their directions of preference differ by 90 degrees.  The
system is surrounded by walls which have doors in the middle through
which the system is closed periodically.  Several flow patterns arise
from these boundary conditions; the most common one is shown
schematically in \figref{crossing}.  In each of the two roads we
observe the formation of lanes as discussed in this section, while the
orientation is different: In one the particles tend to walk on the
right side, in the other on the left.  The region of highest disorder
lies in the center of the system where the two roads meet.  It is
marked with a shaded circle.


\section{Evacuation Simulations}
\label{sec_evac}

In the following we describe results of simulations of a typical
situation, i.e.\ the evacuation of a large room (e.g.\ in the case of
a fire). We try to show the relevance of our model for practical
applications.  It is able to cope with the complex geometrical
structures of any given environment and is therefore of use for
architectural planing and safety estimations of evacuation times of
arbitrary buildings.  First we study a large room without any internal
structure (e.g.\ a ballroom) and just one door.  As an example for a
more complex surrounding we investigate in Sec.~\ref{lechall} the
stylized geometry of a typical lecture hall.  It will be shown that
our model captures the main aspects of the dynamical behaviour of
large crowds even in such a complex situation.

\subsection{Evacuation of a Large Room}
\label{evacuation}

\begin{figure}[ht]
\setlength{\unitlength}{1cm}
\center{
\begin{picture}(14,5)
\put(0.5,0){\scalebox{1}{\includegraphics[width=4cm]{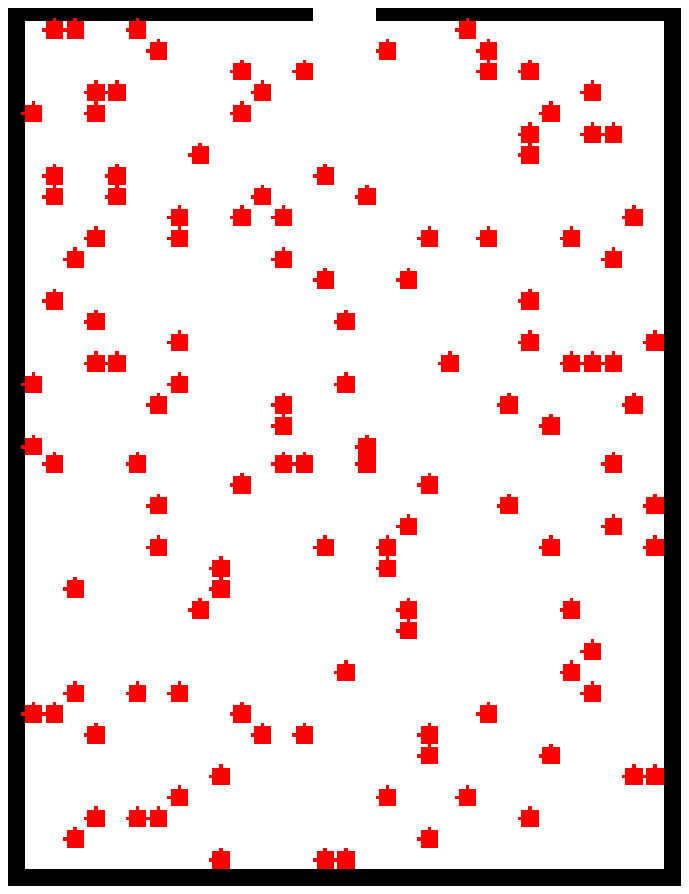}}}
\put(7.5,0){\scalebox{1}{\includegraphics[width=4cm]{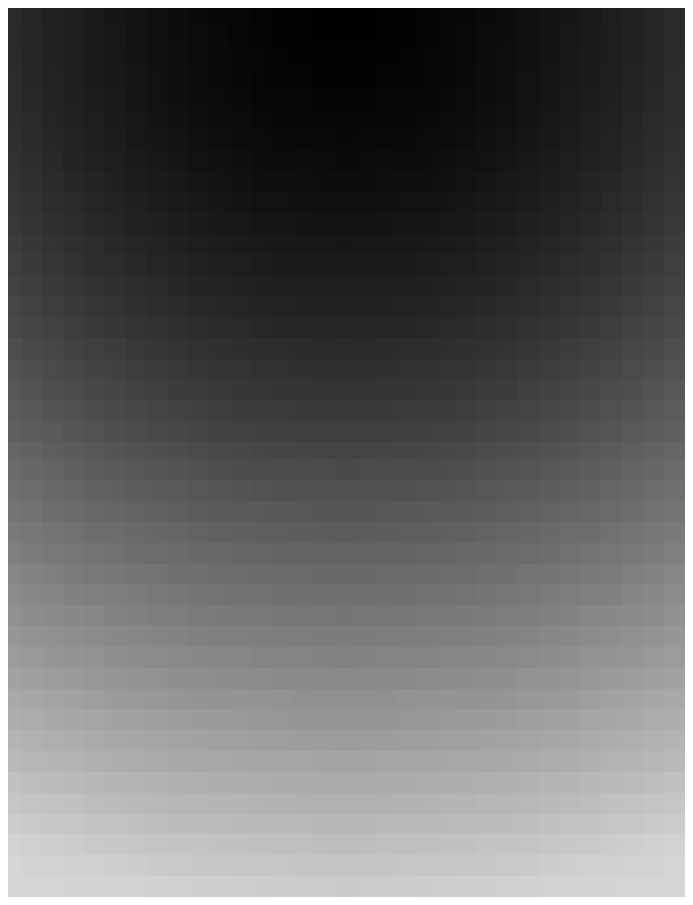}}}
\end{picture}
\caption[]{Ballroom at $t=0$ (left) and static floor field (right).
The strength of the field is proportional to its darkness.}
\label{fig:snaps}}
\end{figure}

\Figref{fig:snaps} shows a large room, e.g.\ a ballroom, with
one door only. 
We have simulated the behaviour of more than 100 people,
which are initially distributed randomly,
leaving this room. It is assumed that they have no 
knowledge of the exact location of the exit. The only information they
get is through the floor fields. 
The static floor field has been chosen such that its strength
decreases radially from a maximum value at the door to zero at the 
corners opposite to the door. 
This is already sufficient to achieve a complete evacuation of the
room. The measured evacuation times depend on the coupling parameters
to the static and dynamic floor fields and the parameters specifying
the evolution of the dynamic floor field.  Without any fine-tuning of
these parameters it is already possible to find reasonable evacuation
times.

\begin{figure}[ht]
\setlength{\unitlength}{1cm}
\center{
\begin{picture}(14,5)
\put(0.5,0){\scalebox{1}{\includegraphics[width=4cm]{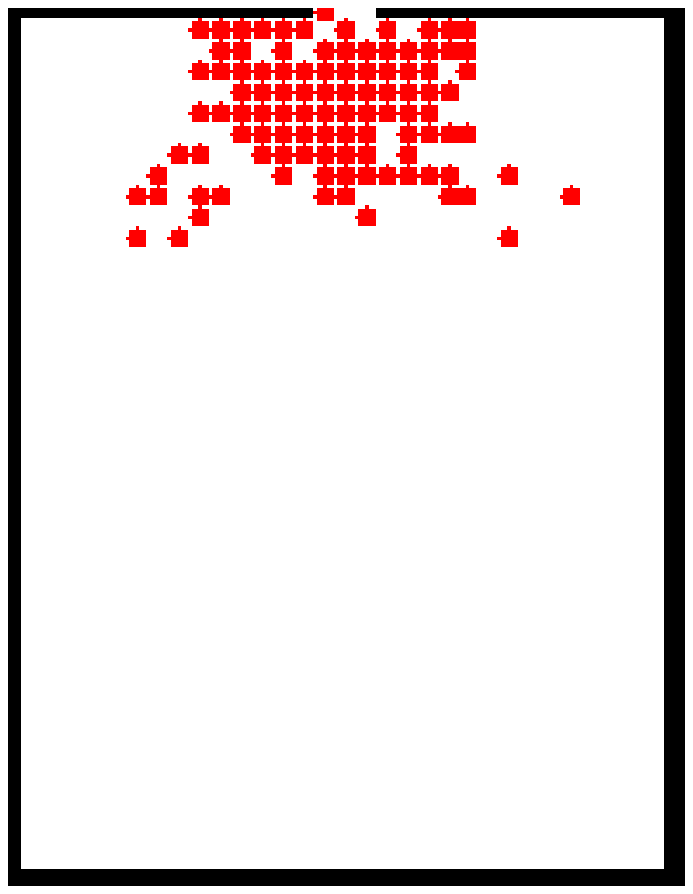}}}
\put(7.5,0){\scalebox{1}{\includegraphics[width=4cm]{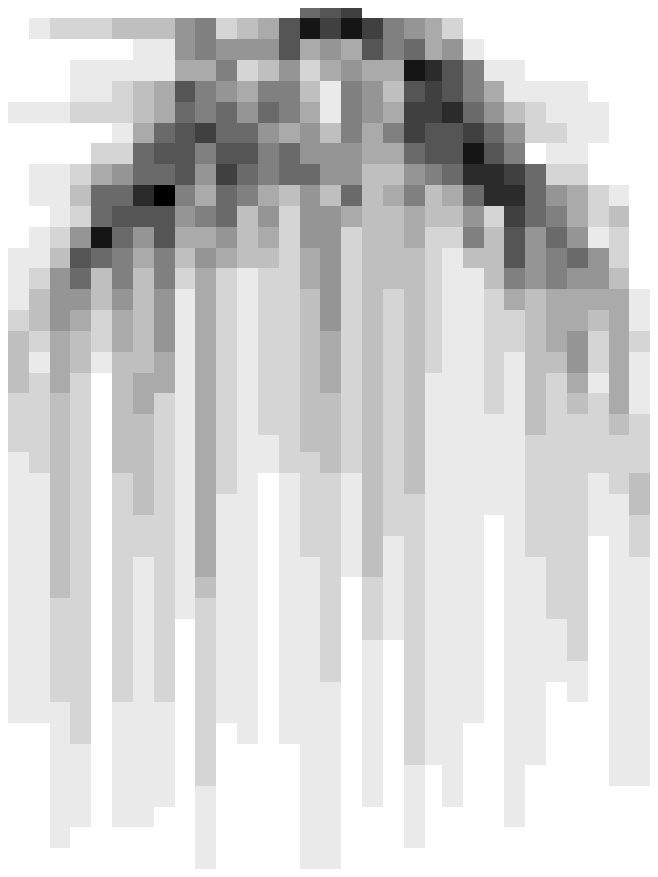}}}
\end{picture}
\caption[]{Evacuation of a large room: Typical configuration at
an intermediate time (left) and the dynamical floor field created
by the people leaving the room (right).}
\label{fig:snaps2}}
\end{figure}

\Figref{fig:snaps2} shows the situation in the room after several
timesteps of the simulation. The left picture shows the people
crowding in front of the door in the typical half-circle way. The
right picture depicts the underlying dynamical field, i.e.\ the
virtual trace left by the pedestrians.  The main traces in the
direction of the door are easily spotted.

One interesting result of these simulations concerns the
influence of the attractive interaction of the pedestrians.
We have found that fluctuations in the measured evacuation times
become much more dominant if the coupling to the static field is small
\cite{ourpaper}. In this case they mainly follow the trace of other
people in the hope that they know the way towards the exit.
This corresponds to a situation where the exit can not be seen, e.g.\ 
due to failing lights or if the room is smoke-filled.
This shows that for the interpretation of evacuation simulations
just studying average evacuation times might lead to wrong conclusions.

\subsection{Evacuation in Complex Geometries: Lecture Hall}
\label{lechall}

\begin{figure}[ht]
\setlength{\unitlength}{1cm}
\center{
\begin{picture}(14,5)
\put(0.5,0){\scalebox{1}{\includegraphics[width=4.5cm]{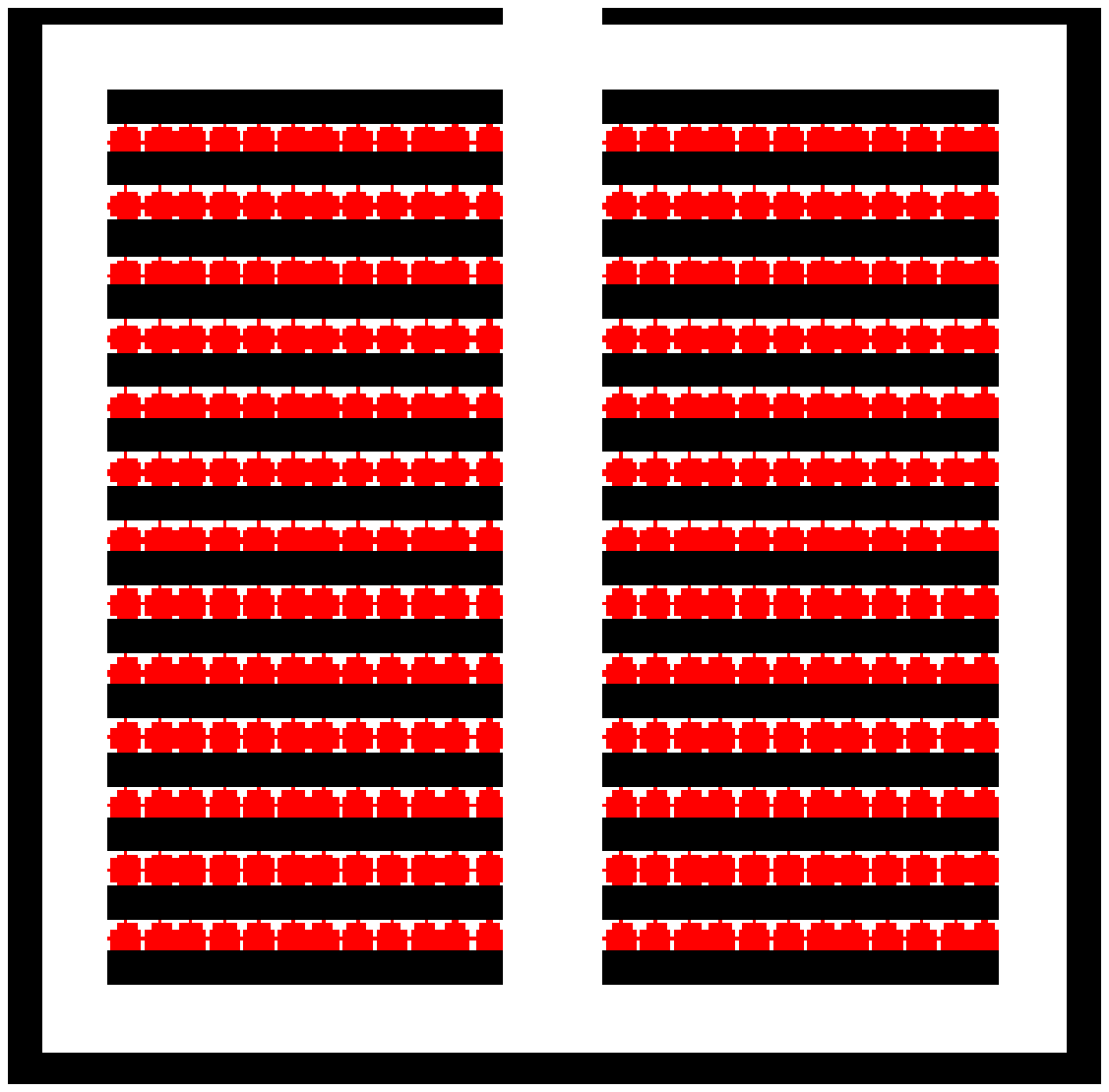}}}
\put(7.5,0){\scalebox{1}{\includegraphics[width=4.5cm]{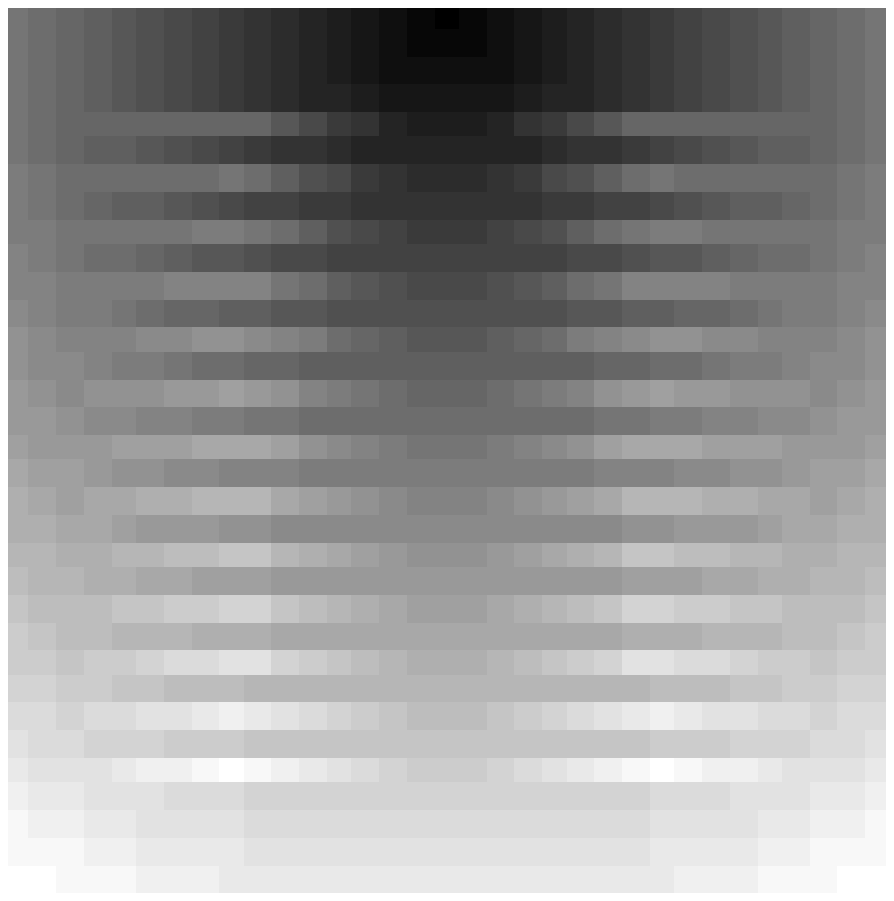}}}
\end{picture}
\caption[]{Lecture hall at $t=0$ (left) and the static floor field
used in the simulations (right).}
\label{pic1}}
\end{figure}

The typical feature of a lecture hall is its compartimentation through
large tiers, which act as non-traversable obstacles for the audience.
Starting point for the simulations at $t=0$ is a fully occupied
lecture hall, i.e.\ all avaliable seats are taken (see \figref{pic1}).
The audience then tries to leave the hall through a single door at the
front of the room.  As in the simpler example of Sec.~\ref{evacuation}
the main guidance in the direction of the door is accomplished by the
static floor field described above.  The rather complicated
geometrical structure of the situation involves some sophisticated
problems in the definition of this field. On the one hand, people have
to orientate themselves into the direction of the door.  On the other
hand, they have first to leave the tiers (which is always possible in
two directions). Without going into detail one can say that these
problems can be solved by the superposition of different suitable
static fields. The resulting field is shown in \figref{pic1}.

\begin{figure}[ht]
\setlength{\unitlength}{1cm}
\center{
\begin{picture}(14,5)
\put(0.5,0){\scalebox{1}{\includegraphics[width=4.5cm]{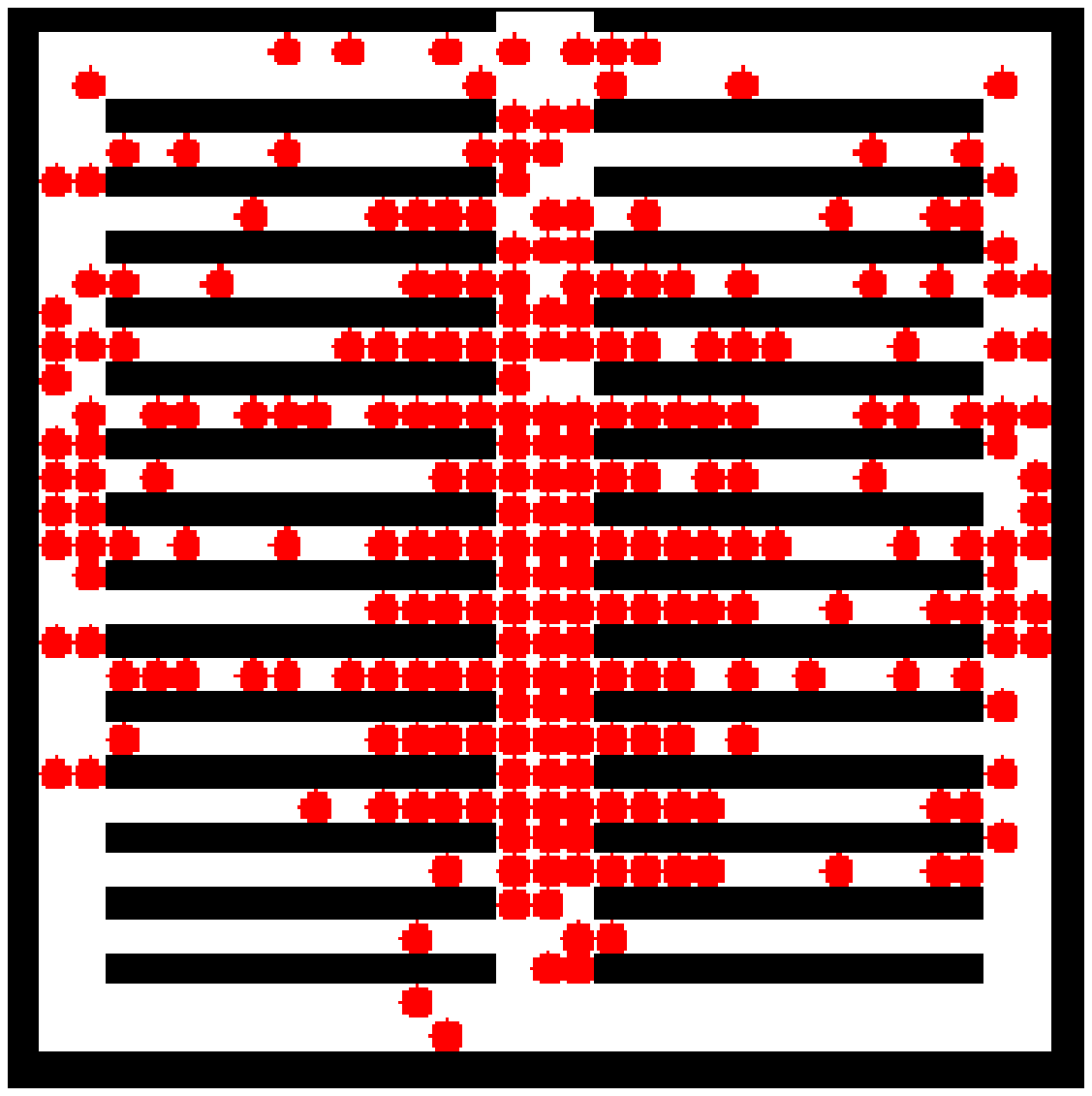}}}
\put(7.5,0){\scalebox{1}{\includegraphics[width=4.5cm]{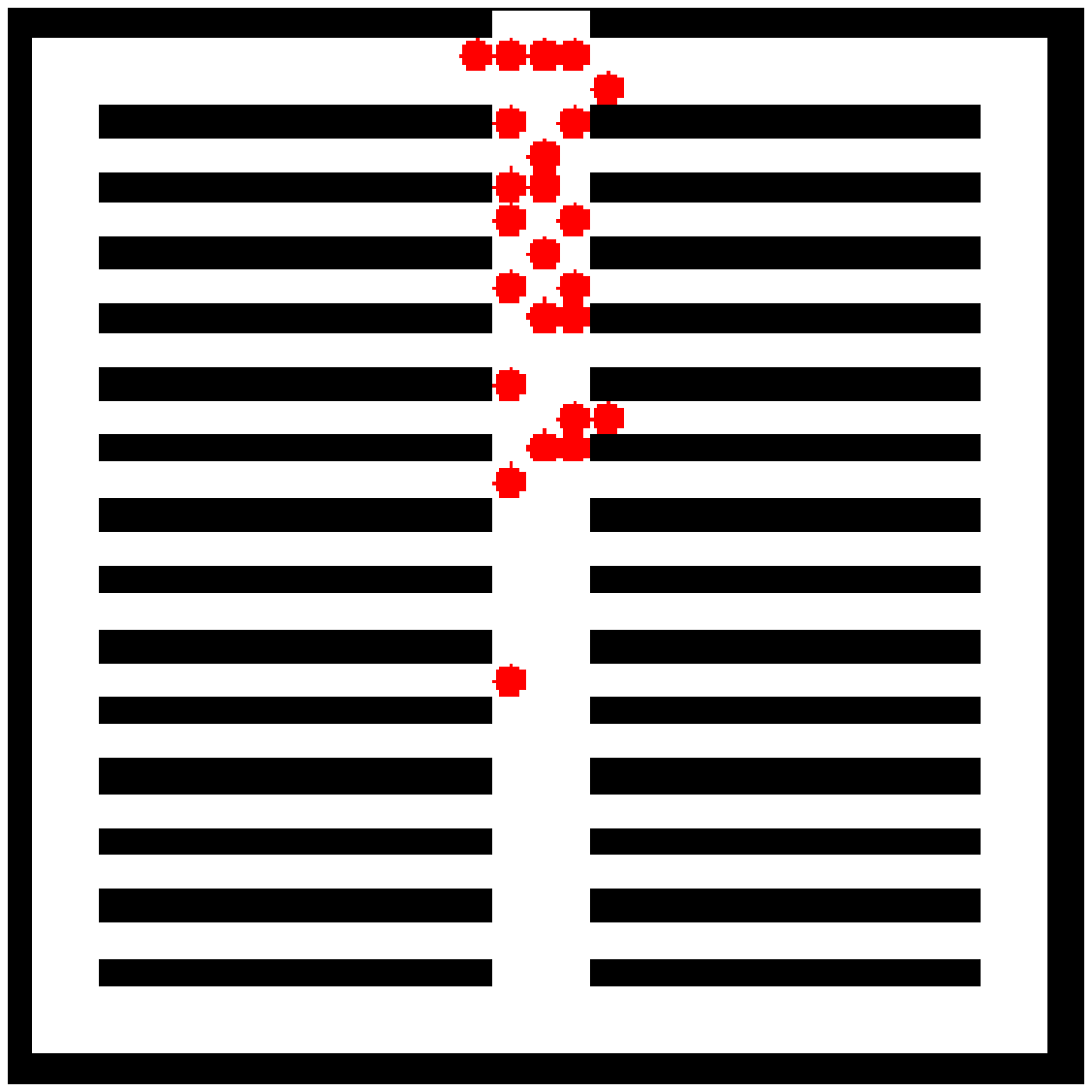}}}
\end{picture}
\caption[]{Typical configurations at later times.}
\label{pic2}}
\end{figure}

The size of the lattice is $33\times 33$ cells with $312$ particles on
it.  \Figref{pic2} shows typical stages of the dynamics of the model
(the left picture displays the hall after $134$ and the right picture
after $590$ timesteps).  In this example we have neglected the
coupling to the dynamical floor field and therefore the whole
evacuation process is driven solely by the static field.

As one can see in these pictures the model is able to describe such a
complex evacuation procedure in a very realistic way. The majority
tries to leave the hall through the main corridor in the middle of the
room.  Therefore it is very crowded and the typical jammed motion is
easy to spot. However some people try to reach the front door by
detouring through the smaller corridors at the sides of the hall, as
in reality.

The whole evacuation process takes about $650$ timesteps which 
corresponds to approximately $3$ minutes of real time. This seems to 
be a rather realistic result which has been obtained without any
fine-tuning of parameter values.

This simple example already shows that our model is able to describe 
rather complex behaviour in complicated geometries. Therefore it can play 
its part in the field of the managemet of evacuation processes.
An example for this will be given in the next section.

\subsection{Optimization of Evacuation Times}
\label{risk}

\begin{figure}[ht]
\setlength{\unitlength}{1cm}
\center{
\begin{picture}(14,5)
\put(0.5,0){\scalebox{1}{\includegraphics[width=4.5cm]{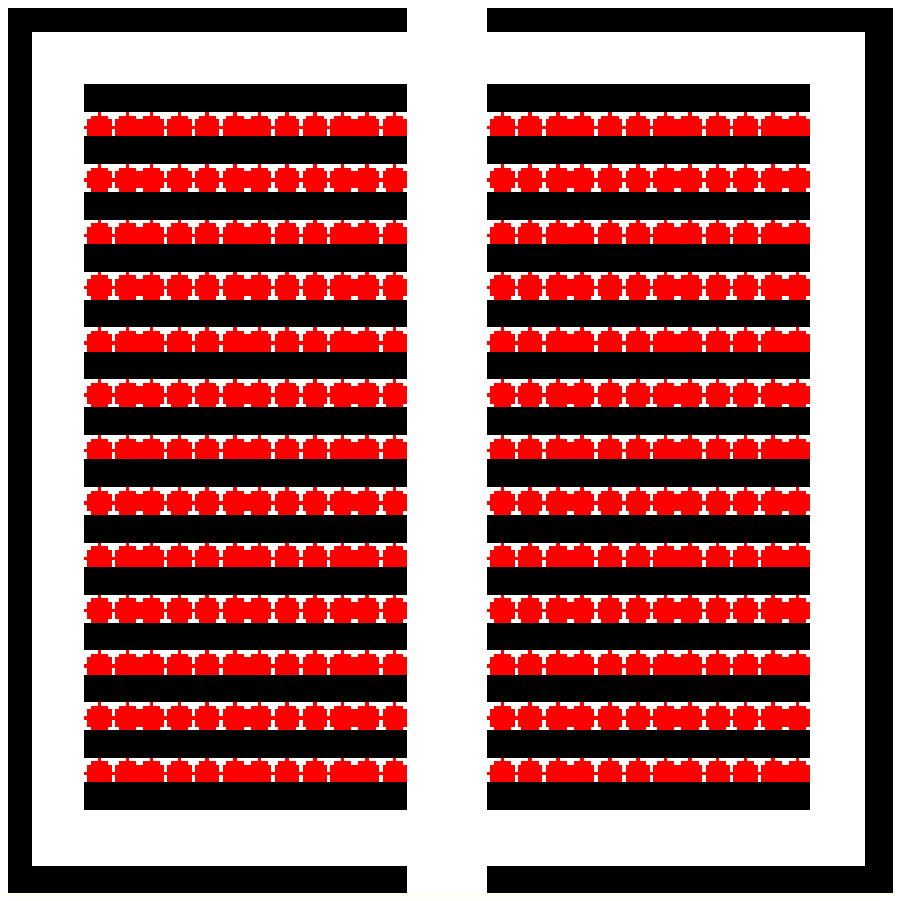}}}
\put(7.5,0){\scalebox{1}{\includegraphics[width=4.5cm]{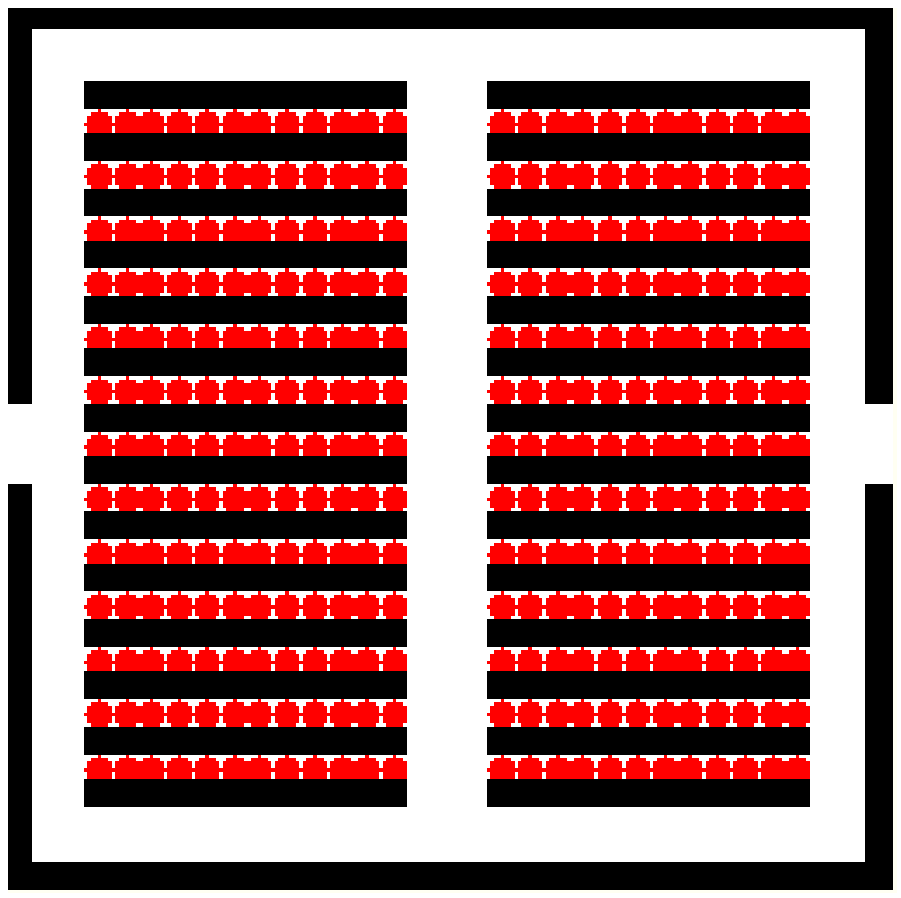}}}
\end{picture}
\caption[]{The two floor planes for hall $A$ (left) and hall $B$ (right).}
\label{pic3}}
\end{figure}

In this subsection we present a simple application of our model in the
field of risk analysis for the design of large buildings. Due to the
high speed of a single simulation one is able to average over a large
number of different evacuation scenarios (which are realised through
different sets of random numbers). Therefore it is possible to make
predictions about average evacuation times (and its fluctuations!) of
different building geometries.  As an example for such measurements we
choose a slightly different situation then presented in the last
subsection. Starting point is again a stylized fully occupied lecture
hall, but now with two doors. In the first floor plane the doors are
located at the front and the bottom of the hall (this geometry will be
named "hall $A$").  In the second floor plane they are placed at the
right and left side ("hall $B$"). These two different geometries are
pictured in \Figref{pic3}.
We have measured the evacuation times for the $312$ people of the
fully occupied rooms. The values for the mean evacuation times $T$ and the 
corresponding variances $\sigma$ are listed in the table below
(in update steps).
\begin{center}
\begin{tabular}{>{$}c<{$}|>{$}c<{$}|>{$}c<{$}}
 & \text{hall}\; A &
  \text{hall}\; B\\ \hline 
  T &     
 560 & 363  \\
  \sigma &   
 85 & 24  
\end{tabular}
\end{center}
Though this result seems to be rather obvious through common sense,
this example shows nevertheless that our model and its rich variations
are very well suited to contribute to investigations in the field of
risk analysis of evacuation processes.


\section{Conclusions}

We have presented simple examples for applications of a new two-dimensional
cellular automaton model for pedestrian dynamics. In contrast to other
CA approaches the model is able to reproduce the collective effects which
are characteristic for pedestrian motion.

We can determine the complete statistical properties (including
variances) of observables like evacuation times.  This knowledge is of
major importance if one wants to establish risk management techniques
that are nowadays used for the hedging of financial assets all over
the world \cite{bp}.

Due to the high computational speed (simulations can be run 10 to 100
times faster than real time) our model is applicable to time-critical
tasks like emergency management.




\begin{thebibliography}{99}

\bibitem{part1} A.\ Schadschneider: {\em Cellular Automaton Approach to 
Pedestrian Dynamics - Theory}, these proceedings

\bibitem{ourpaper}
C.\ Burstedde, K.\ Klauck, A.\ Schadschneider, J.\ Zittartz:
accepted for publication in Physica {\bf A}

\bibitem{diplom} 
C.\ Burstedde: Diploma Thesis, Universit\"at zu K\"oln (2001); available
for download at {\tt http://www.burstedde.de/carsten/diplom.html}

\bibitem{panic} D.\ Helbing, I.\ Farkas, T.\ Vicsek: 
Nature {\bf 407}, 487 (2000)

\bibitem{social} D.\ Helbing, P.\ Molnar: Phys.\ Rev.\ {\bf E51}, 4282 (1995)

\bibitem{bp}
J.-P.\ Bouchaud and M.\ Potters: {\em Theory of Financial Risk}
(Cambridge University Press 2000)

\end{thebibliography}
\end{document}